\documentclass[12pt,amsmath,amssymb,nofootinbib,superscriptaddress]{revtex4-1}
\usepackage{amsmath} %
\usepackage{pstricks}
\usepackage{color} 
\usepackage{amssymb}
\usepackage{slashed}
\usepackage{graphicx,amsfonts,layout,appendix,subfigure}
\input{epsf} 

\usepackage{verbatim} 
\usepackage{upgreek}
\usepackage{float} 
\usepackage{microtype}
\def\gev{\,\text{GeV}^2}
\def\beq{\begin{equation}}
\def\eeq{\end{equation}}
\def\bea{\begin{eqnarray}}
\def\eea{\end{eqnarray}}
\def\beqn{\begin{eqnarray}} 
\def\eeqn{\end{eqnarray}}

\def\nn{\nonumber}

\def\ln#1{\mathrm{ln}\left(#1\right)}

\newcommand\alphas{\alpha_{\mathrm{S}}}

\def\beq{\begin{equation}} \def\eeq{\end{equation}}
\def\beqn{\begin{eqnarray}} \def\eeqn{\end{eqnarray}}
 
\def\nn{\nonumber}

\begin{document} 

\begin{titlepage}
\renewcommand{\thefootnote}{\fnsymbol{footnote}}
\begin{flushright}
     \end{flushright}
\par \vspace{10mm}

\begin{center}
{\Large \bf
The polarized photon distribution function}
\end{center}

\par \vspace{2mm}
\begin{center}
{\bf Daniel de Florian}~$^{(a)}$\footnote{{\tt deflo@unsam.edu.ar}}, {\bf Lucas Palma Conte}~$^{(a)}$\footnote{{\tt lpalmaconte@unsam.edu.ar}},{\bf Gabriel Fernando Volonnino}~$^{(a)}$\footnote{{\tt gabriel.volonnino@gmail.com}}

\vspace{5mm}

${}^{(a)}$
International Center for Advanced Studies (ICAS), ICIFI \& ECyT-UNSAM, 25 de Mayo y Francia,
(1650) San Mart\'\i n,  Pcia. Buenos Aires, Argentina

\vspace{5mm}

\end{center}

\par \vspace{2mm}
\begin{center} {\large \bf Abstract} \end{center}
\begin{quote}
\pretolerance 10000
We employ the LuxQED approach to compute the polarized photon PDF (photon pPDF). This approach expresses the pPDF in terms of the structure functions $g_1$ and $g_2$. Different models for the structure functions are employed according to the parameter space region. The resulting pPDF is approximately of the order of $x$ times the unpolarized PDF. The relative uncertainty in the photon pPDF reaches up to $50\%$ for $x \sim 10^{-3}$, decreasing to approximately $10\%$ for higher values of $x$. The computation of the photon pPDF will be essential for improving the precision of polarized calculations and will have fundamental implications for studies at the future Electron-Ion Collider (EIC).
\end{quote}
\begin{flushleft}

\end{flushleft}
\end{titlepage}

\setcounter{footnote}{0}


\setcounter{footnote}{0}
\renewcommand{\thefootnote}{\fnsymbol{footnote}}

\section{Introduction}
The parton distribution functions (PDFs) are a fundamental piece of a hard scattering process, they represent, at least at the leading order, the probability of finding a parton with a momentum fraction $x$ in a hadron target. With the substantial increase in the accuracy of experimental measurements over the last few decades, it has become imperative to enhance the precision of theoretical calculations. To achieve this goal, precise fittings for PDFs are indispensable. In nearly all studied processes, such as $pp$ scattering or $ep$ scattering, the most important PDFs required are those of quarks and gluons within a proton target. Nonetheless, the photon PDF is emerging as a crucial distribution for incorporating Quantum ElectroDynamic (QED) effects and enhancing the precision of our calculations.
Several global analyses have attempted to fit the photon PDF from data\cite{Martin:2004dh,Ball:2013hta, Schmidt:2015zda, Harland-Lang:2016kog}, however, they led to results with significant uncertainties.
In that context, the LuxQED work \cite{Manohar:2016nzj,Manohar:2017eqh} proposed a method to calculate the photon PDF that avoids the use of phenomenological models and enhances the accuracy of the calculation. The main feature of this method is that it allows us to express the photon PDF in terms of the proton's structure functions.

In the present work, our focus is on the polarized case, with the aim of understanding how spin polarization in protons arises from their constituent structure. To comprehend the contribution of each constituent, analogous to the unpolarized case, we study the polarized parton distribution functions (pPDFs). These are defined as the difference between the PDFs of partons with positive helicity and those with negative helicity within a proton of positive helicity, i.e., $\Delta f = f^+ - f^-$.
Several fits to polarized PDFs have been proposed, including those by DSSV \cite{deFlorian:2014yva, DeFlorian:2019xxt}, NNPDF \cite{Nocera:2014gqa}, JAM collaboration \cite{Ethier:2017zbq} and  MAP collaboration \cite{Bertone:2024taw}. These studies have arrived at important conclusions, such as the non-negligible contribution of $\Delta g$ to the proton spin. However, due to the relatively limited availability of experimental data compared to the unpolarized case, these fits only provide constraints for a narrow range of proton momentum fractions. In general, uncertainties are large, particularly at low-$x$.
In this context, the future Electron-Ion Collider (EIC), which will enable a much wider kinematic range and achieve unprecedented precision for polarized measurements \cite{AbdulKhalek:2021gbh,Accardi:2012qut}, is expected to provide new insights into the spin share of the proton in terms of its fundamental components. Considering the forthcoming measurements, it is essential to increase the precision of the theoretical calculations. In this sense, adding QED effects to the perturbative calculations is crucial. Along these lines, in \cite{deFlorian:2023zkc}, we presented the QED corrections to the evolution of the polarized PDFs, and a recent work has computed the single-inclusive production of prompt photons in electron-proton collisions with the same accuracy \cite{Rein:2024fns}. However, in both studies, toy models were used to represent the polarized photon PDF, which, as discussed in these works, is a key density for these corrections. While some papers have conducted approximate calculations of the elastic contribution to the photon pPDF and discussed its significance in the $ep\rightarrow e\gamma X$ process \cite{Gluck:2002fi,Gluck:2002cm}, a comprehensive analysis of all contributions to the pPDF was not performed. For this reason, in the present study, our main goal is to compute the polarized photon PDF, $\Delta\gamma(x,Q^2)$, employing the LuxQED approach \cite{Manohar:2016nzj,Manohar:2017eqh}. As demonstrated in \cite{Manohar:2017eqh}, this approach is also applicable to the polarized case. They provide an expression for $\Delta\gamma$ in terms of the structure functions $g_1$ and $g_2$. Similar to the unpolarized case, computing $\Delta\gamma$ requires knowledge of the polarized structure functions over a wide range of $Q^2$ and $x$, encompassing both elastic and resonance as well as perturbative behaviour.
To model $g_1$ and $g_2$ in each parameter region, we will use both experimental data and phenomenological models. Throughout the paper, we will explore different models and discuss their validity in each region. For the perturbative contribution to the photon pPDF, we will achieve an accuracy of $\mathcal{O}(\alpha \alphas)$.
The outline of the paper is as follows. Section \ref{sec:Photon pPDF} reviews the general features of the LuxQED approach for the computation. In section \ref{sec:structure-input} we discuss the experimental data inputs and the phenomenological models used for our numerical evaluation in each parameter space region. In section \ref{sec:highorder}, we show how to incorporate high-order contributions into the computation. The final results on the photon pPDF are given in section \ref{sec:result}. We present our conclusions in section \ref{sec:conclu}.
\section{Polarized photon parton distribution}
\label{sec:Photon pPDF}

In the LuxQED work \cite{Manohar:2017eqh}, two approaches are presented for deriving the photon parton distribution function (pPDF) in terms of structure functions. In the first approach, a process is proposed in which a Beyond the Standard Model (BSM) probe interacts with Standard Model (SM) particles solely through the exchange of a photon. The methodology involves calculating this BSM process in two different ways: firstly, using the known structure functions; and secondly, using the photon pPDF, where the photon pPDF emerges at leading order. By equating the results from these two calculations, the photon distribution can be derived. It is crucial to verify that the final distribution is independent of the specific parameters of the BSM probe, ensuring the universality of the result. This approach is validated by performing the calculation with two different BSM probes and obtaining consistent results.

The second method defines the (p)PDF as the matrix element of a (p)PDF operator, following the formalism outlined in \cite{Collins:1989gx}. In this approach, the photon (and other partons) distribution functions are defined through the matrix elements of specific PDF operators, grounded in the operator product expansion and factorization theorems commonly used in Quantum Chromodynamics (QCD). These matrix elements can be expressed in terms of the same structure functions used in the first approach, providing a rigorous and general framework for defining the pPDF. A significant advantage of this method is that facilitates the inclusion of higher-order corrections, enhancing the precision of the pPDF.
Therefore, we adopt the second approach because it offers a clearer path to incorporating higher-order QCD and QED corrections, as seen in Sec. \ref{sec:highorder}. 

The photon pPDF in terms of $g_1$ and $g_2$ is given by  (Eq. (6.14b) in \cite{Manohar:2017eqh})),
\beqn
\Delta\gamma(x,\mu^{2})&=&\frac{8\pi}{\alpha(\mu^{2})\left({\cal S}\mu\right)^{2\epsilon}}\frac{1}{\left(4\pi\right)^{D/2}}\frac{1}{\Gamma(D/2-1)}{\bf\times}\nn\\
&&\int_{x}^{1}\frac{\mathrm{d}z}{z}\int_{Q^2_{min}}^{\infty}\frac{\mathrm{d}Q^{2}}{Q^{2}}\alpha_{D}^{2}(q^{2})\left(Q^{2}(1-z)-x^{2}{m_{p}^{2}}\right)^{D/2-2}\times \label{eq:photonDdim}\\
&&\left\{\left(4-2z-\frac{4m_{p}^{2}x^{2}}{Q^{2}}-4\frac{D-4}{D-2}\frac{Q^{2}(1-z)-x^{2}m_{p}^{2}}{Q^{2}}\right)g_{1,D}(x/z,Q^{2})\right.\nonumber\\&&-\;\frac{8m_{p}^{2}x^{2}}{Q^{2}z}\left(1-\frac{D-4}{D-2}\frac{Q^{2}(1-z)-x^{2}m_{p}^{2}}{Q^{2}}\right)g_{2,D}(x/z,Q^{2})\Bigg\}\,,\nn
\eeqn
where $\mu^2$ is the scale at which we evaluate the pPDF, $x$ the momentum fraction, $m_p$ the proton mass, $Q^2_{\min}=m_p^2\,x^2/(1-z)$, $D=4-2\epsilon$, $\mathcal{S}^2=e^{\gamma_E}/(4\pi)$, $g_{i,D}$ with $i=1,2$, are the structure functions in $D$ dimension, $\alpha(\mu^2)$ is the QED running coupling at scale $\mu^2$ and, lastly, we define,
\beqn
\alpha_D(Q^2)=\frac{\alpha(\mu^2) (\mu {\cal S})^{2 \epsilon}}{1-\Pi_D(\mu^2,Q^2)},
\label{eq:qedcoupling}
\eeqn
where $\Pi_D(\mu^2,Q^2)$ represents the QED vacuum polarization function in $D$ dimensions along with its associated $\overline{\mathrm{MS}}$ counterterm. As explained in Sec. 6 of \cite{Manohar:2017eqh}, to solve Eq.(\ref{eq:photonDdim}), it is convenient to split the integral over $Q^2$ into two parts: the ``Physical" term ($m^2_p\, x^2/(1-z)\rightarrow \mu^2 /(1-z)$) and the ``MS-conversion" term ($\mu^2/(1-z)\rightarrow \infty$), the expression for the photon pPDF is
\beqn
\Delta \gamma(x,\mu^2)=\Delta \gamma^{\mathrm{PF}}(x,\mu^2)+\Delta \gamma^{\mathrm{con}}(x,\mu^2).
\label{eq:photon}
\eeqn
The ``Physical" term is given by (Eq. (6.16b) in \cite{Manohar:2017eqh})
\beqn
\Delta\gamma^{\mathrm{PF}}(x,\mu^{2})&=&\displaystyle\frac{1}{2\pi\alpha(\mu^{2})}\int_{x}^{1}\frac{\mathrm{d}z}{z}\int_{Q^2_{min}}^{\frac{\mu^{2}}{1-z}}\frac{dQ^2}{Q^2}\alpha^{2}(Q^{2})\times\nn\\ &&\displaystyle\left\{\left(4-2z-\displaystyle\frac{4m_{p}^{2}x^{2}}{Q^{2}}\right)g_{1}(x/z,Q^{2})-\frac{8m_{p}^{2}x^{2}}{Q^{2}z}g_{2}(x/z,Q^{2})\right\},\label{eq:photonpf}
\eeqn
where can set $D=4$ ($\epsilon\rightarrow0$) since this term does not present divergences.
The ``MS-conversion" term is (Eq. (6.17b) in \cite{Manohar:2017eqh}),
\beqn
\Delta\gamma^{\mathrm{con}}(x,\mu^{2})=\frac{8\pi\,(S\mu)^{-2\epsilon}}{\alpha(\mu^{2})}\frac{1}{(4\pi)^{D/2}}\frac{1}{\Gamma(D/2-1)}\int_{x}^{1}\frac{\mathrm{d}z}{z}(1-z)^{D/2-2}\int_{\frac{\mu^{2}}{1-z}}^{\infty}\frac{\mathrm{d}Q^{2}}{Q^{2}}\times\nn\\
(Q^{2})^{D/2-2}\,\alpha_{D}^{2}(Q^{2})\biggl\{\biggl(4-2z+4\frac{\epsilon}{1-\epsilon}(1-z)\biggr)\,g_{1,D}(x/z,Q^{2})\biggr\},
\label{eq:photoncon}
\eeqn
where we eliminate the terms proportional to $\frac{m_p^2}{Q^2}$, since taking $\mu^2$ to be sufficiently large implies $Q^2\gg m_p^2$.
The integral of the ``Physical" term covers the whole low-$Q^2$ region and part of the high-$Q^2$ region. It can be integrated numerically, as we will see, using different models for the structure functions according to the point in the parameter space where it is to be evaluated. On the other hand, the integral of the ``MS-conversion" term covers only the high-$Q^2$ region, therefore, we can use the perturbative approximation for the structure functions. If we restrict our analysis to the lowest order of the calculation, $\mathcal{O}(\alpha)$, we can utilize the leading-order structure function $g_1^{\text{LO}}$, which does not depend on $Q^2$. Also, we can set $\alpha_D(Q^2)\simeq\alpha(\mu^2) (\mu \mathcal{S})^{2\epsilon}$, since the evolution with energy (the vacuum polarization in Eq.(\ref{eq:qedcoupling})) is of higher orders. An important observation is that due to the integral run to infinity, there is a divergence when $\epsilon \rightarrow 0$. This divergence is absorbed by the $\overline{{\mathrm{MS}}}$ counterterm, which we don't show explicitly in Eq.(\ref{eq:photoncon}). Replacing the assumptions discussed above in  Eq.(\ref{eq:photoncon}) and performing the $Q^2$ integral, we obtain (Eq. (6.21b) in \cite{Manohar:2017eqh}), 
\beqn
\Delta\gamma_\text{LO}^{\overline{{\mathrm{MS}}}\,\mathrm{con}}(x,\mu^{2})=\frac{\alpha(\mu^{2})}{2\pi}\int_{x}^{1}\frac{\mathrm{d}z}{z}4(1-z)g^{\mathrm{LO}}_{1}(x/z,\mu^{2})+\text{high orders},
\label{eq:photonconLO}
\eeqn 
where the label $\overline{{\mathrm{MS}}}$ indicates that is computed in the $\overline{{\mathrm{MS}}}$-scheme. The structure function at leading order can be expressed in terms of the quarks pPDFs, as follows,
\beqn
g_1^{\mathrm{LO}}(x_{bj},\mu^2)=\frac{1}{2}\sum_{\{q\}} e_q^2 \, (\Delta q(x_{bj},\mu^2)+\Delta \bar{q}(x_{bj},\mu^2))
\label{eq:g1LO}
\eeqn
where the sums runs over all quark flavours $q=u,d,s,b,t$ and, $e_q$ and $\Delta q$ are the charge and pPDF of each flavour, respectively. 
In Sec. \ref{sec:highorder}, we extend our analysis to order $\mathcal{O}(\alpha \alphas)$.
\section{Computation}
\label{sec:structure-input}
\begin{figure} 
    \centering
    \includegraphics[width=0.8\textwidth]{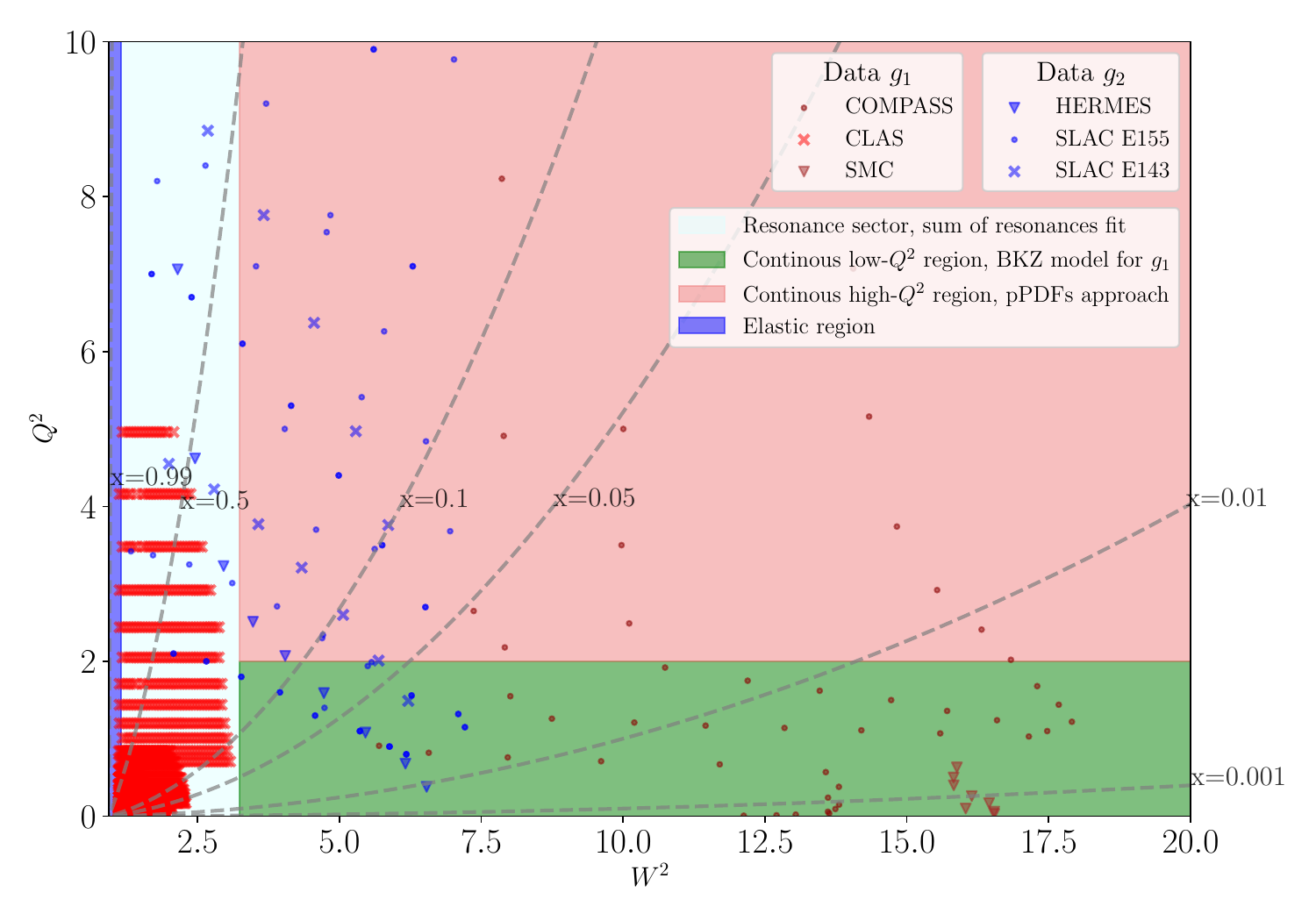}
    \caption{Illustrating each region of the parameter space ($W^2$,$Q^2$), the high-$Q^2$ region in red, the continuous low-$Q^2$ region in dark green, the resonance sector in light green, and the elastic region in blue. The red data points represent $g_1$ measurements from COMPASS \cite{COMPASS:2017hef} (red circles) , CLAS \cite{CLAS:2021apd,CLAS:2006ozz} (red crosses) and SMC \cite{SpinMuon:1999udj,SpinMuon:1998eqa} (red triangles). The blue data points represent $g_2$ measurements from the SLAC E143 collaboration \cite{E143:1998hbs} (blue crosses), the SLAC E155 collaboration \cite{E155:1999eug,E155:2002iec} (blue circles), and HERMES \cite{HERMES:2011xgd} (blue triangles).}
  \label{fig:parameters_space}
\end{figure}

As mentioned above, because the integral of Eq.(\ref{eq:photonpf}) covers a wide range of the parameters $Q^2$ and $x$, we need a complete knowledge of $g_1$ and $g_2$ across this parameter space.
To address this, we will partition the parameter space into three overarching regions, as illustrated in Fig. \ref{fig:parameters_space}. Within each region, we will employ phenomenological models in conjunction with experimental data to delineate the structure functions.
Firstly, we have the elastic region (blue sector in Fig. \ref{fig:parameters_space}), constrained by $W^2<(m_p+m_{\pi^0})^2$, where $m_{\pi^0}$ is the neutral pion mass and $W^2$ is defined as $W^2=m_p^2+\frac{1-x_{bj}}{x_{bj}}Q^2$. Secondly, the resonance sector (light green), where $ (m_p+m_{\pi^0})^2<W^2< W^2_{\mathrm{res}}=3.24\gev$ and the continuous low-$Q^2$ region (dark green), where $W^2>W^2_{\mathrm{res}}$ and $Q^2 < Q^2_{per}=2\gev$. In all these segments, the perturbative approach for $g_1$ and $g_2$ cannot be employed. Lastly, the perturbative sector or continuous high-$Q^2$ region (red), characterized by $Q^2 > Q^2_{per}$ and $W^2>W^2_{\mathrm{res}}$, where the perturbative approximation for the structure function becomes applicable.
In the remainder of this section, we will study in detail each region of the kinematic space. For all regions, we utilize the LO approximation for the evolution of the QED running coupling, with the initial condition $\alpha(\mathrm{Rb})=(137.035 998 995)^{-1}$, where $\mathrm{Rb}$ denotes the mass of a $^{87}$Rubidium atom \cite{Aoyama:2019ryr,Bouchendira:2010es}. 
\subsection{Elastic region}
\label{subsec:elastic}
\begin{figure} 
    \centering
    \includegraphics[width=0.75\textwidth]{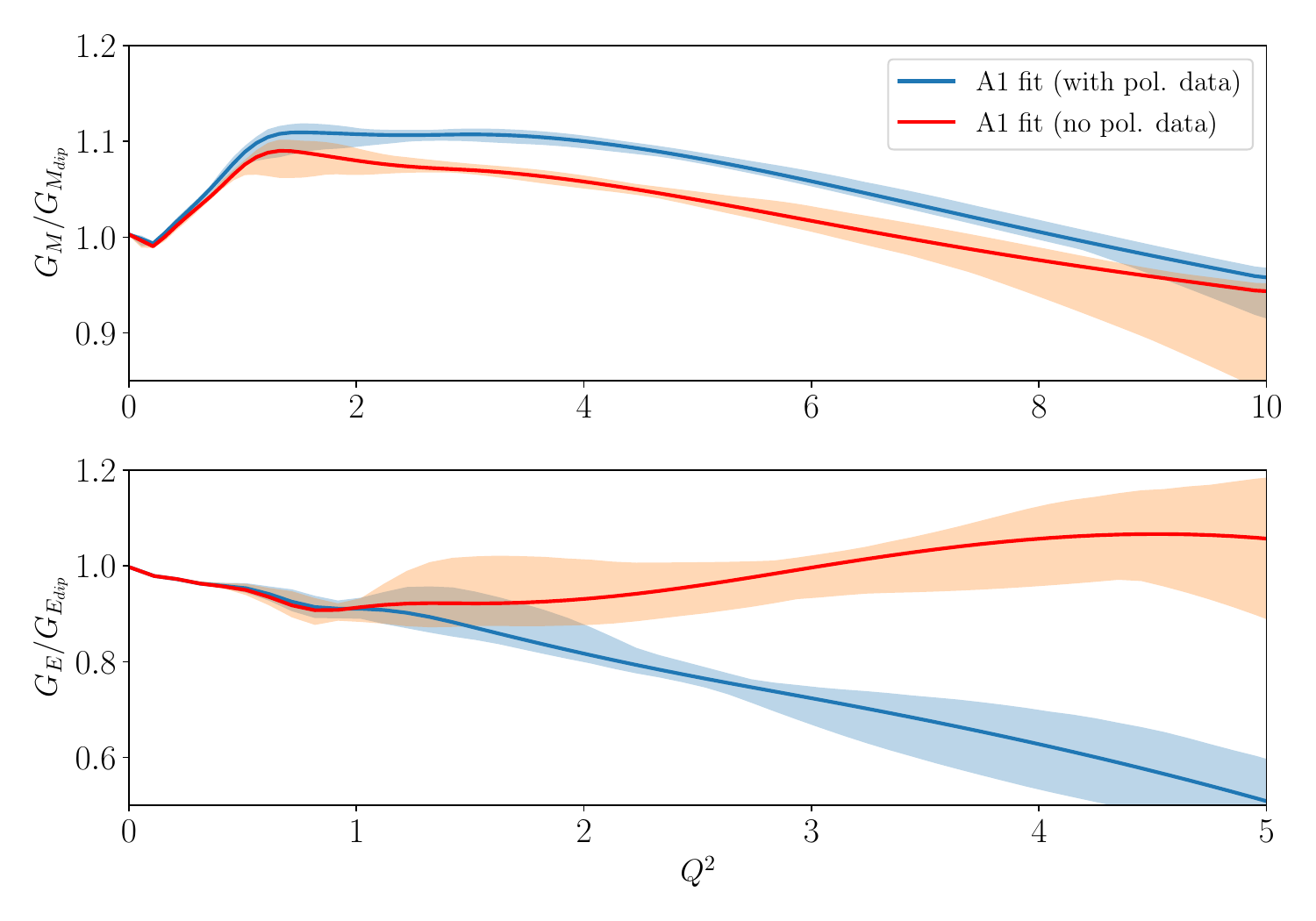}
    \caption{Ratio of the form factors to standard dipole form fitted by the A1 collaboration \cite{A1:2013fsc}.}
    \label{fig:GEGM}
\end{figure}

As we have already mentioned, we define the elastic region as constrained by $W^2<(m_p+m_{\pi^0})^2$. Within this domain, we can express the structure function in terms of the electric and magnetic Sachs form factors of the proton, $G_E$ and $G_M$ respectively \cite{CLAS:2017qga},
\beqn
g_1^{\mathrm{ela}}(x_{bj},Q^2)&=&\frac{1}{2}\frac{G_E(Q^2)\,G_M(Q^2)+\tau G_M(Q^2)}{1+\tau}\delta(x_{bj}-1) \label{eq:g1ela},\\
g_2^{\mathrm{ela}}(x_{bj},Q^2)&=&\frac{\tau}{2}\frac{G_E(Q^2)\,G_M(Q^2)-G_M(Q^2)}{1+\tau}\delta(x_{bj}-1)\label{eq:g2ela},
\eeqn
where $\tau=Q^2/(4m_p^2)$ and $x_{bj}$ is the usual Bjorken scale. Note that the delta functions neglect the photon radiation from the proton. However, most of the radiation is smooth and therefore cancels out in inclusive quantities. Anything that deviates from the $\delta$-function approximation represents a correction to the photon distribution beyond our accuracy.
Replacing Eqs.(\ref{eq:g1ela}, \ref{eq:g2ela}) in the Eq.(\ref{eq:photonpf}) we obtain,
\beqn
\Delta \gamma^{\mathrm{ela}}= \frac{1}{2\pi\alpha(\mu^2)}\int^{\frac{\mu^{2}}{1-z}}_{Q^2_{min}} \frac{d Q^2}{Q^2}\alpha^2(Q^2)\left[\frac{G_E(Q^2)G_M(Q^2)}{1+\tau}(2-2x-\frac{2m_p^2x^2}{Q^2})+\frac{tG_M(Q^2)^2}{1+\tau}(2-\frac{2m_p^2x^2}{Q^2})\right].
\label{eq:photonela}
\eeqn
At this point, we have to propose a model for the $G_{E,M}$ Sachs form factors. A first approximation is the well-know dipole form,
\beqn
G_E(Q^2)=\frac{1}{(1+Q^2/m_{dip}^2)}, \quad G_M(Q^2)=\mu_{p}\,G_E(Q^2),
\eeqn
where $m_{dip}^2 =0.71 \gev$ and $\mu_p \simeq 2.793$ is the anomalous magnetic moment of the proton. For $Q^2 = 0$ this form yields the exact results $G_E (0) = 1$ and $G_M (0) = \mu_p$ , while elsewhere it is an approximation. We will see that the elastic region has a high incidence on the final result of the photon pPDF. For this reason, we improve our accuracy using an experimental fit to the form factors published by the A1 collaboration \cite{A1:2013fsc}. The A1 data is limited to $Q^2<1\gev$, however, the paper also includes fits to global data up to $Q^2 \sim 10\gev$. Beyond $10\gev$ we use the A1 fit evaluated at $Q^2=10\gev$ for both $G_E$ and $G_M$. In Fig. \ref{fig:GEGM} we show the form factor fits normalized to the dipole form. The A1 paper includes two classes of fits, one for just unpolarized data (red), and one that also includes polarized data (blue). We adopted the fit with polarized data as our default approach, as it offers stronger constraints on the two-photon exchange (TPE) contributions and provides a more robust framework for our analysis. In Fig. \ref{fig:GEGM} we also show the experimental error bands, which, when propagated in the computation, provide an estimate of the photon pPDF uncertainty. To perform the propagation, we treated the fit uncertainty of the elastic and magnetic form factors as entirely correlated, as proposed in \cite{Manohar:2017eqh}. We compute $\Delta\gamma^{\mathrm{ela}}$, i.e., Eq.(\ref{eq:photonela}), with $G_{E,M}$ evaluated at the extremes of the error bands. Then, we consider the variation of these computations as the uncertainty in the photon pPDF. 
\subsection{Continuous low-$Q^2$ region and resonance sector}
\label{subsec:lowq}
\begin{figure} 
    \centering
    \includegraphics[width=0.7\textwidth]{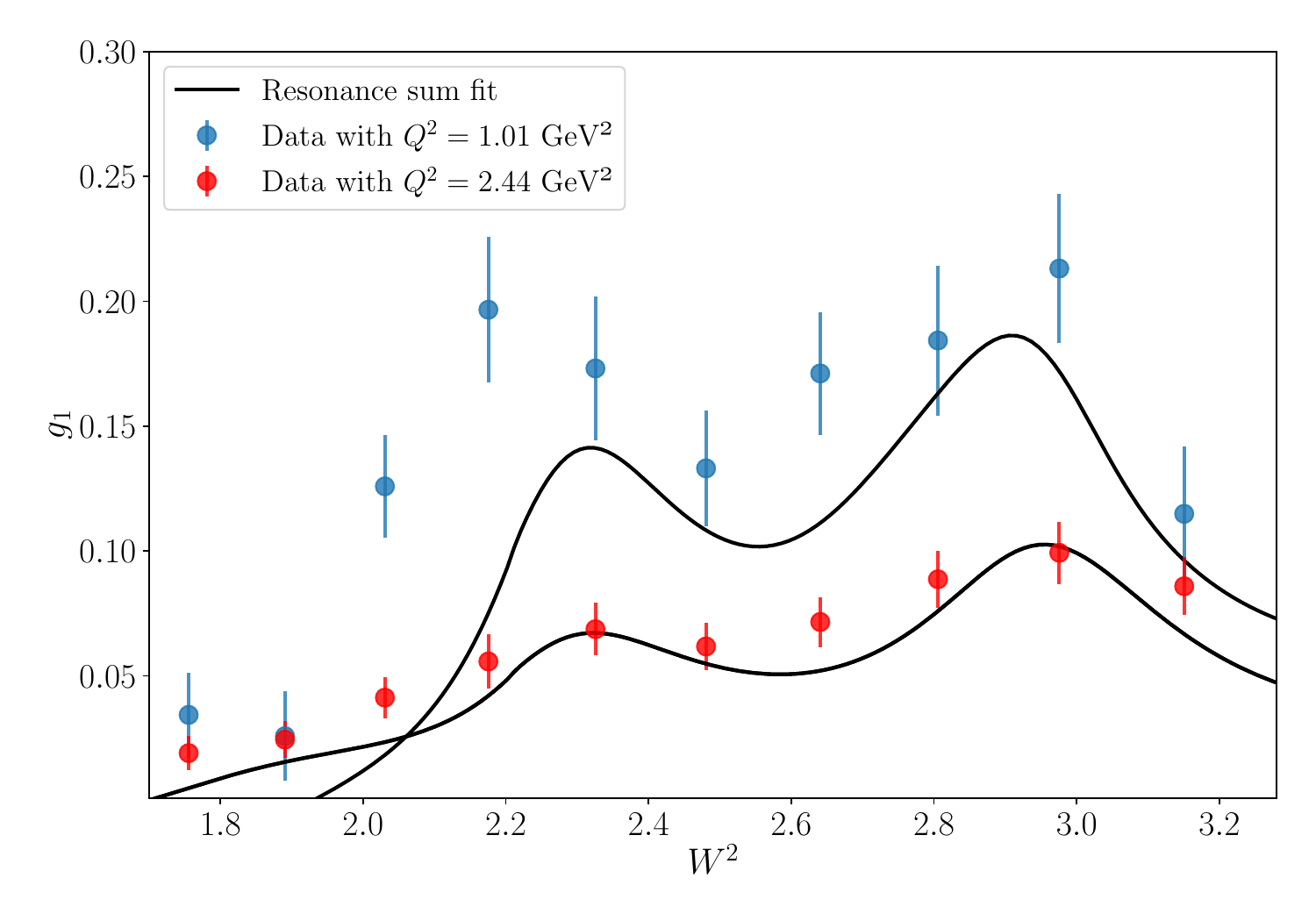}
    \caption{Illustration of a subset of the CLAS data  \cite{CLAS:2021apd} of $g_1$ in the resonance region at $Q^2=1.01 \gev$ (blue circles) and $Q^2=2.44\gev$ (red circles), and the resonances fits of \cite{HillerBlin:2022ltm} (black curves).}
    \label{fig:g1vsx_reonance}
\end{figure}

\begin{figure} 
    \centering
    \includegraphics[width=0.75\textwidth]{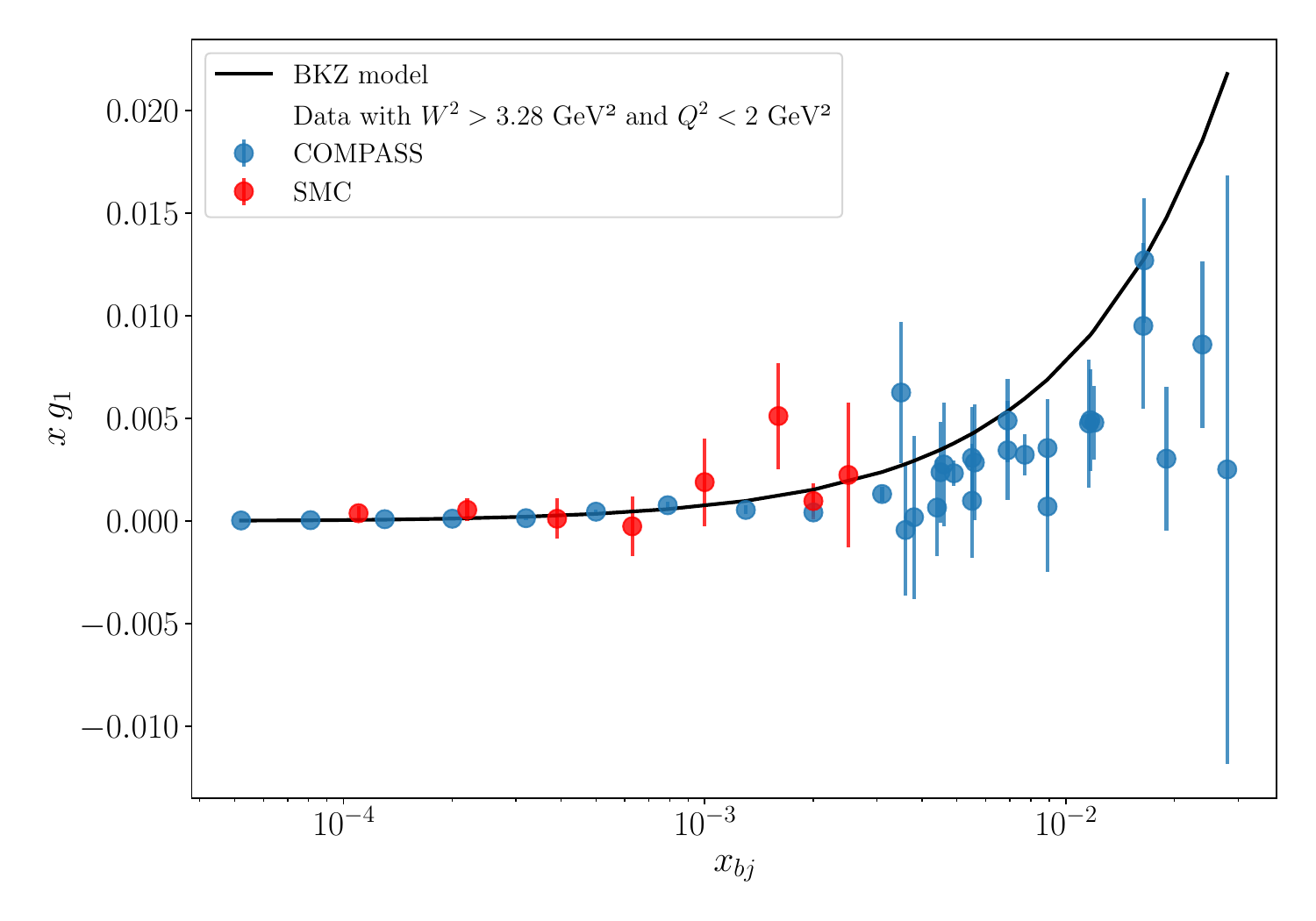}
    \caption{Data for $g_1$ from COMPASS \cite{COMPASS:2017hef,COMPASS:2015mhb} (blue circles) and SMC \cite{SpinMuon:1999udj,SpinMuon:1998eqa} (red circles) in the continuous low-$Q^2$ region. The fit is the BKZ model \cite{Badelek:2003nm,Badelek:2002jr,Badelek_1999} Eq.(\ref{eq:g1bkz}) with the parameter $C$ fitted from the data and the parametrization of the perturbative part using the DSSV pPDF set with QED corrections for the evolution \cite{DeFlorian:2019xxt,deFlorian:2023zkc}.}
    \label{fig:g1vsx_BKZ}
\end{figure}

The continuous low-$Q^2$ region corresponds to $W^2 > (m_p + m_{\pi^0})^2$ and $Q^2 < Q_{\text{per}}^2 $, while the resonance sector corresponds to $(m_p + m_{\pi^0})^2 < W^2 < W_{\text{res}}^2 $. Since the perturbative approximation is inapplicable within these regions, the structure function cannot be computed using that approach. Instead, we rely on directly measured data and phenomenological models.
For the resonance part a study has been carried out which models the resonances of $g_1$ and $g_2$ as a sum of resonance terms \cite{HillerBlin:2022ltm}. They provide a publicly available code enabling the computation of the structure functions for any given values of $Q^2$ and $x$ within the resonance region. The model fitting utilizes data from the CLAS collaboration \cite{CLAS:2006ozz,CLAS:2021apd}, where only longitudinally polarized proton targets are employed, consequently, $g_1$ is inferred from the data by employing a global fit to $g_2$ world data. Additionally, they use data from the RSS collaboration \cite{RSS:2006tbm,ResonanceSpinStructure:2008ceg}, SANE collaboration \cite{SANE:2018pwx}, and the Jefferson Lab Hall A g2p collaboration \cite{JeffersonLabHallAg2p:2022qap}. In these experiments, both $A_{\parallel}$ and $A_{\perp}$ have been determined, enabling the full reconstruction of $g_1$ and $g_2$.
We show the different experiment measurement data points in Fig. \ref{fig:parameters_space}. However, the sum of resonances model is applicable only for $Q^2<7.5 \gev$ \cite{HillerBlin:2022ltm}. For values beyond this threshold, we resort to the perturbative approach outlined in Sec. \ref{subsec:highq}. Nevertheless, this constraint corresponds to high values of $x$, where the structure functions approach zero and the resonances decrease. Therefore, employing the perturbative approach is not expected to significantly alter the final result. In Fig. \ref{fig:g1vsx_reonance}, we show the phenomenological model (black curves) of the structure function $g_1$ as a function of $W^2$ at $Q^2=1.01 \gev$ (blue circles) and $Q^2=2.44\gev$ (red circles), along with the available experimental data for those scales.
Within the continuous low-$Q^2$ region, as shown in Fig. \ref{fig:parameters_space}, there are fewer measurements compared to the resonance region. Therefore, since a direct model from data is not available, we must resort to using phenomenological models for the structure functions. For $g_1$, we employ the BKZ model \cite{Badelek:2003nm,Badelek:2002jr,Badelek_1999}, which is based on Vector Meson Dominance (VMD) concepts \cite{RevModPhys.50.261},
\begin{equation}
\begin{aligned}
g^{\mathrm{bkz}}_1\left(x_{bj}, Q^2\right)
& =C\left[\frac{4}{9}\left(\Delta u_{v a l}^{(0)}(x_{bj})+\Delta \bar{u}^{(0)}(x_{bj})\right)+\frac{1}{9}\left(\Delta d_{v a l}^{(0)}(x_{bj})+\Delta \bar{d}^{(0)}(x_{bj})\right)\right] \frac{M_\rho^4}{\left(Q^2+M_\rho^2\right)^2} \\
& +C\left[\frac{1}{9}\left(2 \Delta \bar{s}^{(0)}(x_{bj})\right)\right] \frac{M_\phi^4}{\left(Q^2+M_\phi^2\right)^2} \\
& +g_1^{\mathrm{per}}\left(x_{bj}, Q^2\right),
\label{eq:g1bkz}
\end{aligned}
\end{equation}
where, $\Delta j^{(0)}$, with $j=u,d,s$ are the pPDF evaluated at $Q^2_0=1\gev$, $M_{\phi}$ and $M_{\rho}$ are the masses of the mesons $\phi$ and $\rho$ respectively, and $g_1^{\mathrm{per}}$ is an extrapolation of the QCD improved parton model structure function to arbitrary values of $Q^2$ and $x$. The parameter $C$ could be computed from DHGHY sum rule \cite{Badelek:2003nm,Ioffe:1994qh}, but instead, we fit that parameter from the experimental data. 
For the parameterization of the perturbative part of Eq.(\ref{eq:g1bkz}), both for $g_1^{\mathrm{per}}$ and the pPDFs, we use the DSSV pPDF set \cite{deFlorian:2014yva,DeFlorian:2019xxt} with the QED corrections presented in \cite{deFlorian:2023zkc}.
In \cite{COMPASS:2017hef}, this model is used in the low-$Q^2$ region, yielding good agreement with experimental data. In Fig. \ref{fig:g1vsx_BKZ}, we depict the experimental data points available for the low-$Q^2$ region, extracted from COMPASS \cite{COMPASS:2017hef,COMPASS:2015mhb} (blue circles) and SMC \cite{SpinMuon:1999udj,SpinMuon:1998eqa} (red circles),  alongside the BKZ model (black curve). It demonstrates good agreement with the experimental data, allowing us to evaluate the $g_1$ function in areas where data are unavailable.
We combine both models using two transition scales, $W^2_{lo}=3\gev$ and $W^2_{hi}=4.5 \gev$, to smooth the transition between resonance and continuous regions, as proposed in \cite{Manohar:2017eqh},
\begin{equation}
g_{1} = 
\begin{cases}
  g_{1}^{\mathrm{res}} & \text{  } W^{2}< W_{\mathrm{lo}}^{2}\,, \\
  (1- \rho(W^{2})) g_{1}^{\mathrm{res}}+\rho(W^{2})g_{1}^{\mathrm{bkz}} & \text{  } W_{\mathrm{lo}}^{2}< W^{2}< W_{\mathrm{hi}}^{2}\,, \\
  g_{1}^{\mathrm{bkz}} & \text{  } W^{2}>\ W_{\mathrm{hi}}^{2}\,,
\end{cases}
\end{equation}
where $g_1^{\mathrm{res}}$ is the phenomenological model from \cite{HillerBlin:2022ltm} and $\rho$ is
\begin{equation}
    \rho{\big(}W^{2}{\big)}\,=\,2\omega^{2}\,-\,\omega^{4}\,,\qquad\quad\omega\,=\,\frac{W^{2}\,-\,W_{\mathrm{lo}}^{2}}{W_{\mathrm{hi}}^{2}-\,W_{\mathrm{lo}}^{2}}\,.
\end{equation}
In this way, we cover the entire resonance and continuous zones at low-$Q^2$ for $g_1$. We use the same smooth transition between the resonance region and the high-$Q^2$ region, replacing $g_{1}^{\mathrm{bkz}}$ with the perturbative approach that we will describe in Sec. \ref{subsec:highq}. On the other hand, the situation for $g_2$ is more complex, as there is no phenomenological model available for the continuous zone. Furthermore, the amount of experimental data for $g_2$ is considerably less than for $g_1$. 
Nevertheless, in Eq.(\ref{eq:photon}), it can be shown that the coefficient multiplying $g_2$, i.e., $r=\frac{8m_p^2 x^2}{z Q^2}$, satisfies $r \leq \frac{2m_p^2}{(W^2_{\text{res}}-m_p^2)}\simeq 0.75$ in the continuous region. Since the coefficient multiplying $g_1$ is much larger ($\sim4$), this constraint allows us to neglect the contribution of $g_2$ in that zone. Note that this is not the case in the resonance region, where $r$ may take larger values, considering that, despite $Q^2$ being low, $x$ is not necessarily small. For the uncertainty associated with $g_1$ and $g_2$, we locate the $N_{\text{data points}}=5$ experimental data points closest to the point where we are evaluating the structure function. Then, we calculate the average experimental uncertainty for that set of measurements. This value is assigned as the uncertainty to the calculated structure function. Subsequently, we propagate it to the photon pPDF. 

\subsection{Continuous High-$Q^2$ region: Perturbative approach}
\label{subsec:highq}
\begin{figure} 
    \centering
    \includegraphics[width=0.75\textwidth]{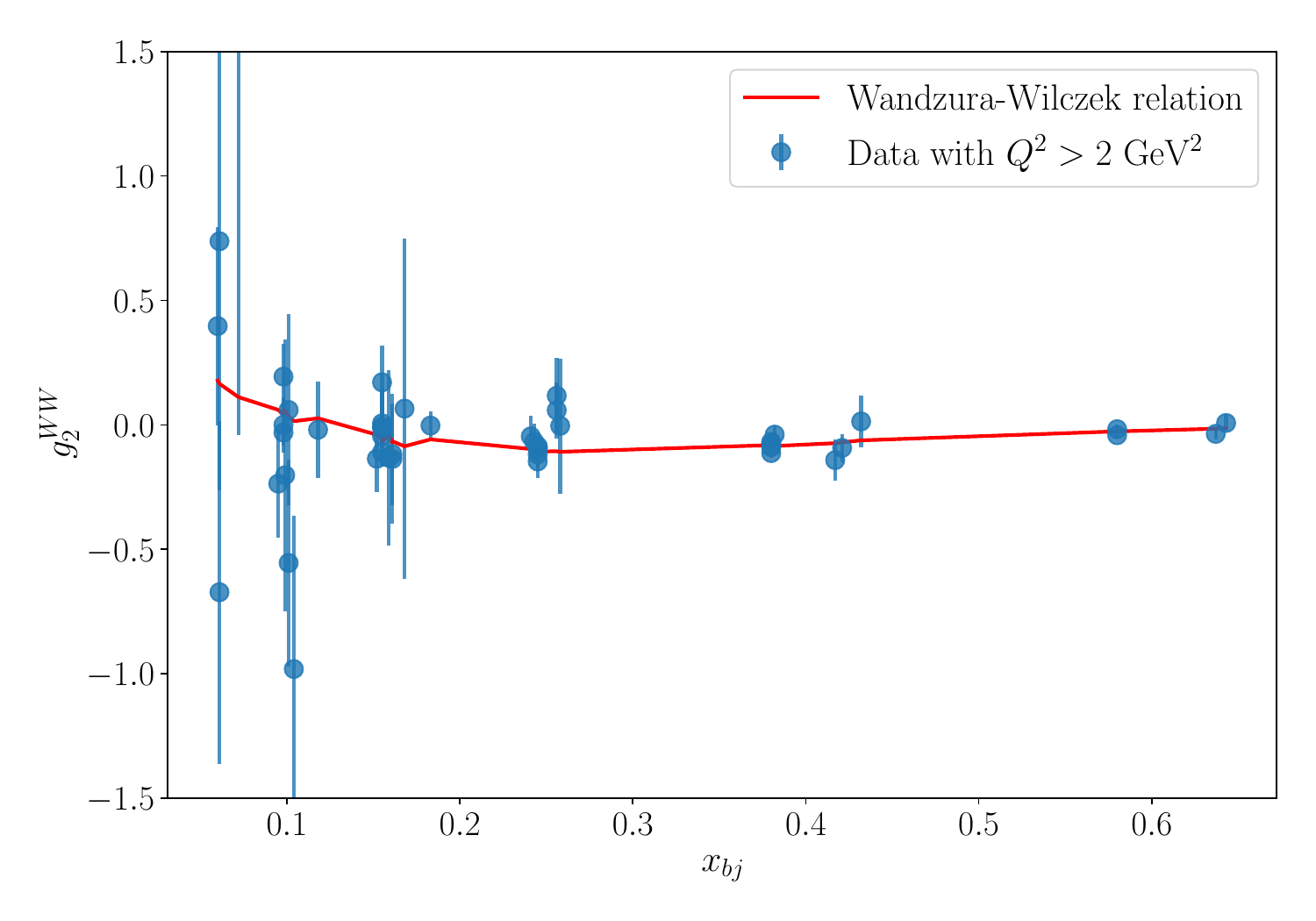}
    \caption{Combined data from SLAC \cite{E143:1998hbs,E155:1999eug} and HERMES \cite{HERMES:2011xgd} for $g_2$ in the high-$Q^2$ region ($Q^2>2\gev$). The fit is the Wandzura-Wilczek model, Eq.(\ref{eq:wandzura-wilzeck}) (red curve).}
    \label{fig:g2_wanzura}
\end{figure}

\begin{figure} 
    \centering
    \includegraphics[width=0.75\textwidth]{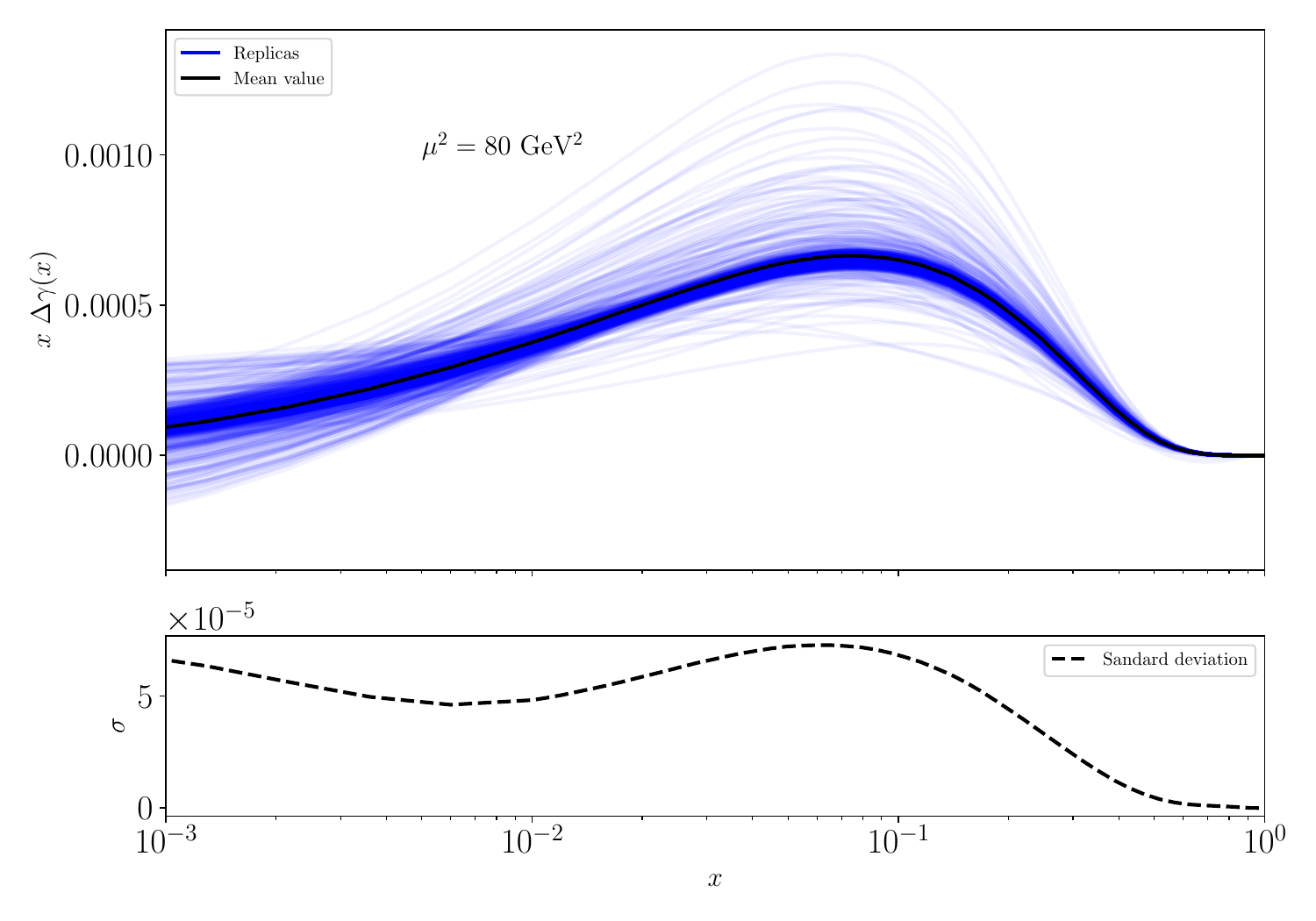}
    \caption{Photon pPDF replicas at $\mu^2=80\gev$ computed with each replica of the $g_1$ from the DSSV pPDF set with QED corrections for the evolution \cite{DeFlorian:2019xxt,deFlorian:2023zkc} (blue curves). Additionally, we show the mean value (black curve, upper plot) and the standard deviation (dashed black curve, lower plot) computed over the ensemble of replicas.}
    \label{fig:photon_replicas}
\end{figure}

In the high-$Q^2$ region, where $Q^2>Q_{\text{per}}^2$ and $W^2>W^2_{\text{res}}$, the perturbative approach is applicable for computing the structure functions. As shown above for the ``MS conversion" we use the LO approximation for $g_1$, i.e., Eq.(\ref{eq:g1LO}). On the other hand, for $g_1$ in the ``Physical" term, Eq.(\ref{eq:photonpf}), we employ the QCD+QED calculation described in \cite{deFlorian:2023zkc}, using the DSSV pPDF set evolved with QED corrections. Certainly, for computing $g_1$, we initially require the photon pPDF. As a first approximation, we employ a toy model $\Delta\gamma=x \, \gamma$ \cite{deFlorian:2023zkc}, using $\gamma$ from the NNPDF set \cite{Bertone:2017bme}. Since the contribution of $\Delta \gamma$ to $g_1$ is small, within the order we are working, it does not significantly affect the final result when calculating the photon pPDF with Eq.(\ref{eq:photon}). In addition, for $g_2$, we apply the Wandzura-Wilczek relation \cite{Wandzura:1977qf}, i.e.,
\begin{equation}
g_{2}^{\mathrm{WW}}\left(x_{bj}\right)=-g_{1}\left(x_{bj}\right)+\int_{x_{bj}}^{1} \frac{d y}{y} g_{1}(y).
\label{eq:wandzura-wilzeck}
\end{equation}
Given that it is an approximation of twist-2, we expect that this relation works well at high-$Q^2$. In Fig.\ref{fig:g2_wanzura}, we illustrate the combination of data from SLAC \cite{E143:1998hbs,E155:1999eug,E155:2002iec} and HERMES \cite{HERMES:2011xgd} with the constraint $Q^2>2 \gev$ (blue circles) and the Wandzura-Wilczek relation (red curve). We can appreciate that the model satisfactorily approximates the data. It will allow us to evaluate $g_2$ in areas where there is no available data.
The high-$Q^2$ region has several uncertainty sources that we have to estimate. Firstly, in the ``Physical" term $\Delta\gamma^{\mathrm{PF}}$, i.e., Eq.(\ref{eq:photonpf}), there is a theoretical uncertainty obtained by independently varying the renormalization $\mu_R$ and factorization $\mu_F$ scales in the calculation of $g_1$. We compute the photon pPDF by varying both non-physical scales in $g_1$ by a factor of 2, i.e., $Q/2<\mu_{F,R}< 2Q$. We take the sum in quadrature of both variations as the uncertainty in the photon pPDF. It should be noted that this uncertainty cannot be estimated for the LO ``MS-conversion" term, Eq.(\ref{eq:photonconLO}), because at the leading order, $g_1$ does not have a dependence on a non-physical scale.
Another source of uncertainty arises from the arbitrary election of the split in the two terms of Eq.(\ref{eq:photon}). The final pPDF should not depend on the scale where we split the integral. Therefore, we can change that scale at $M_Z^2/(1-z)$, with the final result expected to remain independent of that choice, i.e., the missing high-order terms must cancel that variation, a similar concept of standard scale variation. The new expression for the ``Physical" and ``MS-conversion" terms are,
\beqn
\Delta\gamma^{\mathrm{PF}}(x,\mu^{2},M_Z)&=&\displaystyle\frac{1}{2\pi\alpha(\mu^{2})}\int_{x}^{1}\frac{\mathrm{d}z}{z}\int_{Q^2_{min}}^{\frac{M_Z^{2}}{(1-z)}}\frac{dQ^2}{Q^2}\alpha^{2}(Q^{2})\times\nn\\ &&\displaystyle\left\{\left(4-2z-\displaystyle\frac{4m_{p}^{2}x^{2}}{Q^{2}}\right)g_{1}(x/z,Q^{2})-\frac{8m_{p}^{2}x^{2}}{Q^{2}z}g_{2}(x/z,Q^{2})\right\},\\
\Delta\gamma_{\mathrm{LO}}^{\overline{{\mathrm{MS}}}\,\mathrm{con}}(x,\mu^{2},M_Z)&=&\Delta\gamma_{\text{LO}}^{\overline{{\mathrm{MS}}}\,\mathrm{con}}(x,\mu^{2})+\frac{1}{2\pi\alpha(\mu^{2})}\int_{x}^{1}\frac{\mathrm{d}z}{z}\int_{\frac{M_Z^2}{(1-Z)}}^{\frac{\mu^{2}}{(1-z)}}\frac{dQ^2}{Q^2}\alpha^{2}(Q^{2})\nn\times\\ &&\left\{\left(4-2z\right)g^{\text{LO}}_{1}(x/z,Q^{2})\right\}.
\eeqn
By varying $\mu/2<M_Z<2 \mu$, we obtain a photon pPDF with associated uncertainty. 
Finally, we consider the uncertainty in $g_1$ originating from the various PDF replicas within the DSSV set \cite{DeFlorian:2019xxt}. We compute $g_1$ and subsequently derive the photon pPDF using each replica. We take the mean of the photon replicas and the standard deviation as the uncertainty. The upper plot of Fig. \ref{fig:photon_replicas} illustrates the $N_{rep}=999$ replicas of the photon pPDF (blue curves), along with the mean value (black curve). Additionally, the plot below shows the standard deviation computed over the ensemble of replicas (dashed black curve). 
\section{High orders}
\label{sec:highorder}
Referring back to Eq.(\ref{eq:photoncon}), we can improve the accuracy of the ``MS-conversion" part of the photon pPDF computation by incorporating higher orders of $g_1$. The structure function can be expressed in terms of the pPDFs as follows
\beqn
g_{1,D}(x,Q^{2})=\int^1_{x}\frac{dz}{z}\sum_{f\in\{q,l,g,\gamma\}}\mathcal{G}_{1,f}(z,Q^{2},\mu^{2},\epsilon)\,\Delta f(x/z,\mu^{2}),
\label{eq:g1Ddim}
\eeqn
where the sum runs over all partons $f$, $\Delta f(x/z,\mu^{2})$ is the pPDF of each parton. In turn, the coefficients $\mathcal{G}_{1,f}$ can be expressed as a perturbative series in the couplings of QCD and QED,
\beqn
\mathcal{G}_{1,f}=\sum_{i,j=0}^{\infty}\,\left(\frac{\alpha_{s}}{2\pi}\right)^{i}\,\left(\frac{\alpha}{2\pi}\right)^{j}\,\mathcal{G}_{1,f}^{(i,j)}\,,
\label{eq:g1coef}
\eeqn
where $\alpha_s$ is the QCD running coupling. Replacing the various perturbative orders of $g_1$ in Eq.(\ref{eq:photoncon}), we can obtain higher orders of the photon pPDF. It is convenient to define,
\beqn
\Delta\gamma^{\mathrm{con}}(x,\mu^{2})=\sum_{i,j=0}^{\infty}\,\left(\frac{\alpha_{s}}{2\pi}\right)^{i}\,\left(\frac{\alpha}{2\pi}\right)^{j}\,\Delta \gamma^{(i,j)}\,,
\label{eq:photonserie}
\eeqn
where we can obtain $ \Delta \gamma^{(i,j)}$, using $g_1$ at the corresponding order. In the lowest order, the only non-vanishing coefficient is,
\beqn
\mathcal{G}_{1,q}^{(0,0)}&=&e_q^2\delta(1-z),
\eeqn
where $e_q$ is the charge of the quark $q$.
Replacing in Eq.(\ref{eq:g1coef}) and in Eq.(\ref{eq:g1Ddim}) we obtain the structure function at LO given by Eq.(\ref{eq:g1LO}). Consequently, the lowest order of the photon pPDF will yield the same result as Eq.(\ref{eq:photonconLO}). The QCD corrections are given by
\beqn
\mathcal{G}_{f=q,g}^{(1,0)}=\left(\frac{\mu^{2}}{Q^{2}}\right)^{\epsilon}\left[-\,\frac{1}{\epsilon} \Delta B_{1,f=q,g}^{(1,0)}(z)+\Delta C_{1,f=q,g}^{(1,0)}(z)-\epsilon \Delta a_{1,f=q,g}^{(1,0)}(z)\right]+\left[\frac{1}{\epsilon}\Delta B_{1,a}^{(1,0)}(z)\right]_{\mathrm{c.t.}},
\label{eq:wilsonDim}
\eeqn
where the last term is the $\overline{{\mathrm{MS}}}$ counterterm and the functions $\Delta B_{1,f=q,g}^{(1,0)}$, $\Delta C_{1,f=q,g}^{(1,0)}$ and $\Delta a_{1,f=q,g}^{(1,0)}$ can be extracted from \cite{Zijlstra:1993sh,Manohar:2017eqh}.
By substituting Eq.(\ref{eq:wilsonDim}) into Eq.(\ref{eq:g1coef}) and then into Eq.(\ref{eq:g1Ddim}), we obtain $g_{1,D}$. Then, utilizing the "MS-conversion" term, i.e., Eq.(\ref{eq:photoncon}), and performing the $Q^2$ integration, we obtain,
\beqn
\Delta\gamma^{(1,1)}&&=\sum_{f\in\{q,g\}} \int^1_x \frac{dz}{z} \int^1_{x'} \frac{dz'}{z'} \frac{1}{12}\bigg\{\bigg[2 \pi^2-24+24 z-z\pi^2\nn\\
&&+24(z-1)\ln{1-z}+6(z-2)\ln{1-z}^2\bigg]\Delta B_{1,f}^{(1,0)}(z')\label{eq:photonNLO}\\
&& \bigg[12(z-2)\bigg] \Delta a_{1,f}^{(1,0)}(z')+\bigg[24(1-z)-12(z-2)\ln{1-z}\bigg]\Delta C_{1,f}^{(1,0)}(z')\bigg\}\Delta f(x'/z',\mu^2),\nn
\eeqn
where $x'=x/z$. We only keep the finite term; the infinite term is absorbed in the counterterm for the photon pPDF. We numerically solved the convolutions of Eq. (\ref{eq:photonNLO}) and found that the relative corrections to the total photon pPDF are approximately $\sim 5\%$. We could repeat this procedure to achieve higher orders by using the perturbation expansion for $\alpha_D$ and for the structure functions in $D$ dimensions. However, because of the large uncertainty, as seen in Sec. \ref{sec:result}, we consider that it is not necessary to go beyond $\mathcal{O}(\alpha \alphas)$.
\section{Results}
\label{sec:result}
\begin{figure} 
    \centering
    \includegraphics[width=0.75\textwidth]{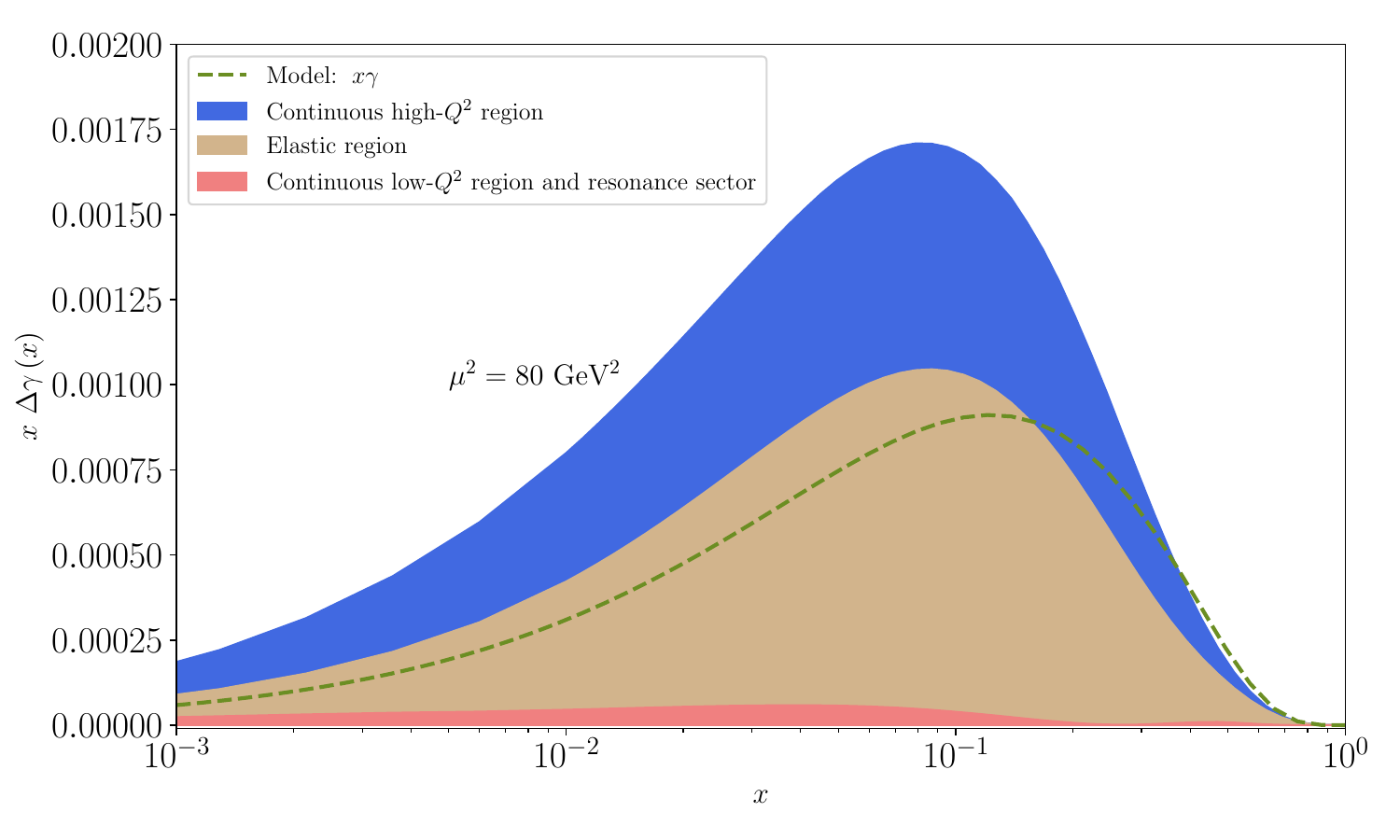}
    \caption{Results for the photon pPDF at $\mu^2=80 \gev$. We show the contribution of the high-$Q^2$ region (blue), the elastic region (brown) and the low-$Q^2$ region and resonance sector (red). The dashed green curve is the toy model $x\gamma$.}
    \label{fig:photon_contribuciones}
\end{figure}

\begin{figure} 
    \centering
    \includegraphics[width=0.75\textwidth]{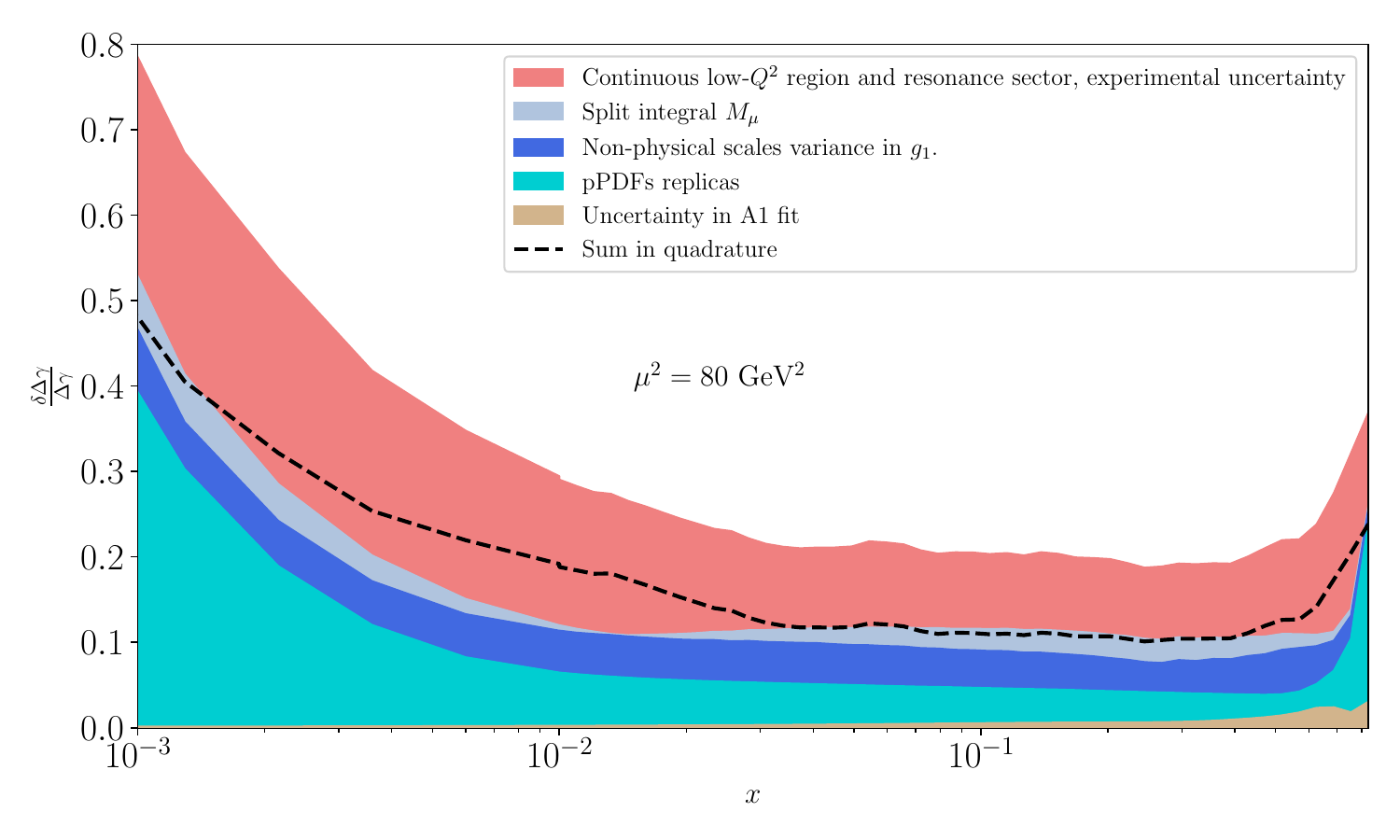}
    \caption{Relative uncertainty for the photon pPDF at $\mu^2=80 \gev$. The blue range indicates the uncertainty stemming from the high-$Q^2$ region, encompassing uncertainties from the replicas of pPDFs of the DSSV set, as well as variations in the non-physical scales and the split scale $M_Z$ (Sec. \ref{subsec:highq}). In red, the uncertainty originates from the low-$Q^2$ region, where an experimental error is incorporated into the phenomenological models BKZ and the resonance model from \cite{HillerBlin:2022ltm} (Sec. \ref{subsec:lowq}). Finally, the brown segment represents the elastic uncertainty arising from the form factors A1 fit \cite{A1:2013fsc} (Sec. \ref{subsec:elastic}). The dashed black curve is the sum in quadrature of all errors.}
    \label{fig:photon_errores}
\end{figure}

\begin{figure}
    \centering
    \includegraphics[width=0.8\textwidth]{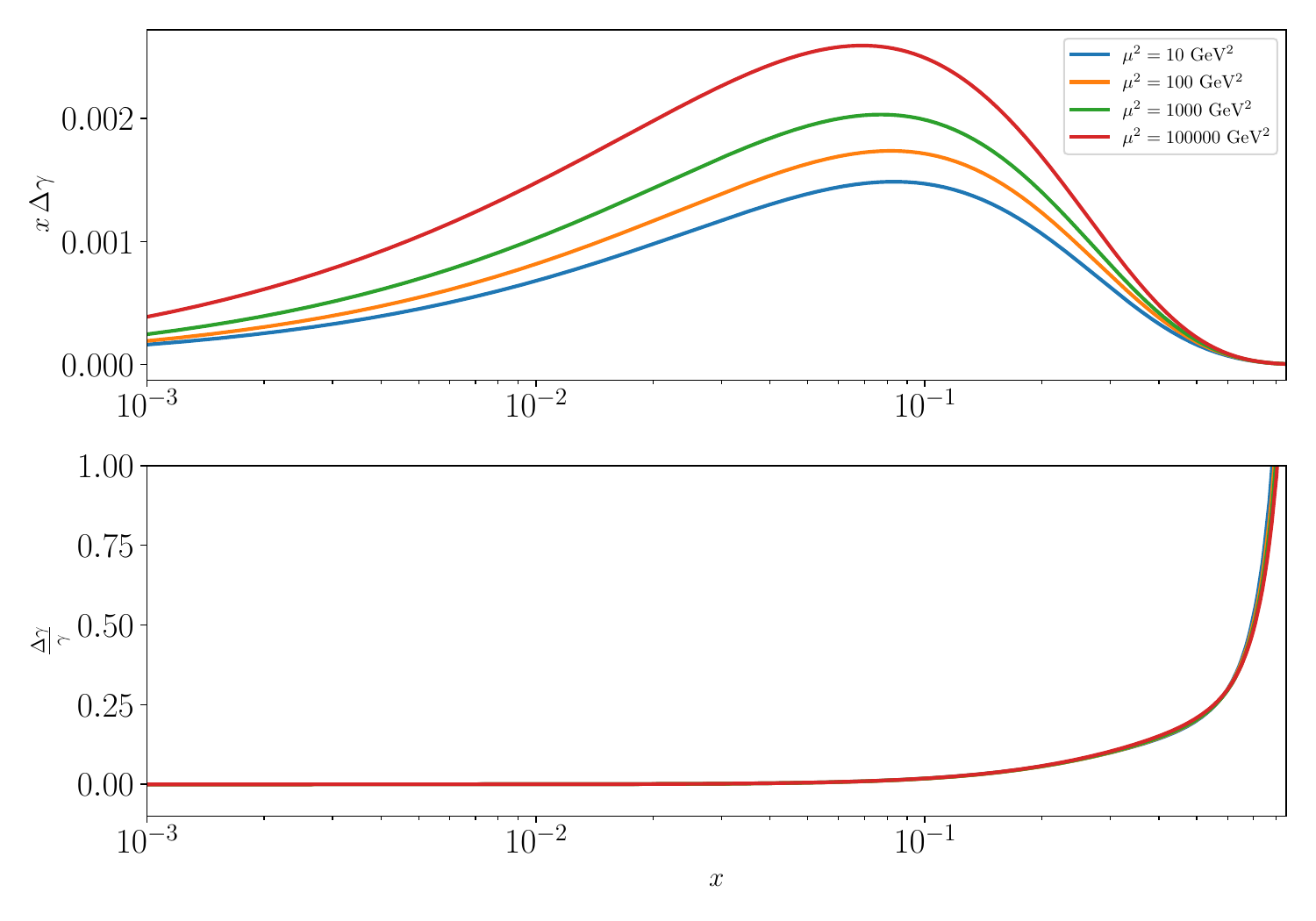}
    \caption{Photon pPDF evolved, with mix-order corrections to the DGLAP equations, to different scales of $\mu^2$ (upper plot). Ratio between the polarized and the unpolarized case, with the unpolarized case given by the NNPDF set \cite{Bertone:2017bme} (lower plot).}
    \label{fig:polvsunpol}
\end{figure}

In Fig. \ref{fig:photon_contribuciones} we show all the contributions to the photon pPDF described in Sec. \ref{sec:structure-input}, at $\mu^2=80\gev$. We can see that the largest contribution comes from the elastic region (brown), followed by the high-$Q^2$ region (blue) and lastly the low-$Q^2$ and resonance region (red). We also show the toy model used in \cite{deFlorian:2023zkc} for the photon pPDF $\Delta \gamma=x\,\gamma$ (green dashed curve), where for the unpolarized photon PDF we use the NNPDF set \cite{Bertone:2017bme}. It can be observed that the prediction of that model is roughly of the same order as the newly computed photon pPDF in this work.
In Fig. \ref{fig:photon_errores}, we present all the relative uncertainties discussed in Sec. \ref{sec:structure-input}. The uncertainty in the resonance and low-$Q^2$ region, illustrated in red, originates from the experimental error in that region added to the phenomenological model, as explained in Sec. \ref{subsec:lowq}. The range of blue colours indicates uncertainties arising from the high-$Q^2$ region, as described in Sec. \ref{subsec:highq}. Firstly, starting from the top, there is uncertainty originating from the split in the integral; secondly, the variation of the unphysical scales $\mu_F$ and $\mu_R$ in $g_1$; and lastly, the error from the standard deviation of the DSSV set replicas. Finally, the brown area indicates the uncertainty originating from the elastic region, as discussed in Sec.\ref{subsec:elastic}, these uncertainties arise from the experimental error in the A1 fit.
The dashed black curve represents the quadratic sum of the errors. It can be seen that for $x\sim 10^{-3}$, the relative uncertainty reaches approximately $50\%$, and decreases to approximately $10\%$ for $x\sim 10^{-1}$. 
The two primary sources of uncertainty arise from the pPDF replicas and experimental errors in the low-$Q^2$ and resonance regions.
In the upper plot of Figure \ref{fig:polvsunpol}, we display the photon pPDF evolved to different scales $\mu$ using QCD+QED DGLAP equations, as solved in \cite{deFlorian:2023zkc}. The changes resulting from the evolution are small.
Additionally, in the lower plot, we illustrate the ratio between the polarized PDF and the unpolarized case, $\frac{\Delta \gamma}{\gamma}$, using the NNPDF set for the latter \cite{Bertone:2017bme}. As expected, this quotient is less than $1$, and as $x\rightarrow1$, it approaches $1$. Lastly, taking the first moment of our result, it gives the contribution of the photon to the proton spin. We present results for the truncated moments since we cannot determine the complete first moment due to high uncertainties at low-$x$. At an scale $\mu^2=10 \gev$ we obtain, $\int^1_{0.001} \Delta \gamma\, dx \simeq 0.0049 \pm 0.0008$. This is a small value in comparison with the quark and gluon contribution, $\int^1_{0} \frac{1}{2}\Delta \Sigma+\Delta g\, dx$, which already saturates the proton spin sum rule \cite{Borsa:2024mss}.
\section{Conclusions}
\label{sec:conclu}
We calculate the PDF of the polarized photon using the approach introduced in the LuxQED paper \cite{Manohar:2017eqh}, employing various models for the structure functions depending on the region of the parameter space analysed.
For the elastic region, we employ an expression for the structure functions in terms of the Sachs form factors, measured by the A1 collaboration. In both the continuous low-$Q^2$ region and resonance sector, we employ a combination of experimental data and phenomenological models. Specifically, for the resonance region, we adopt a model for both $g_1$ and $g_2$ based on resonance sums as presented in \cite{HillerBlin:2022ltm}. For the continuous part, we employ the BKZ model for $g_1$ proposed in \cite{Badelek:2003nm,Badelek:2002jr,Badelek_1999}, which is grounded on VMD concepts. Also, we demonstrate that the contribution of $g_2$ in this region is negligible. Lastly, in the high-$Q^2$ region, we use the PDF approach, employing the DSSV pPDF set with QED corrections to describe $g_1$. For $g_2$, we apply the Wandzura-Wilczek relation.
The computed polarized photon PDF is of the order of $x \gamma$, which is the model we proposed in \cite{deFlorian:2023zkc}. Additionally, we estimated the uncertainty associated with the calculation to be approximately $50\%$ for $x\sim 10^{-3}$, decreasing to around $10\%$ for larger $x$ values. We computed the truncated first moment of the photon, at $\mu^2=10 \gev$ we found, $\int^1_{0.001} \Delta \gamma\, dx \simeq 0.0049 \pm 0.0008$. Taking into account future measurements from the EIC, the photon pPDF will be crucial for improving the precision of polarized calculations. Recently, the work by \cite{Rein:2024fns} discusses the importance of the photon pPDF in the computation of semi-inclusive production of prompt photons. It would be interesting to test our pPDF calculation in this kind of process.
\begin{acknowledgements}
This work is partially supported by CONICET and ANPCyT.
\end{acknowledgements}
%

\end{document}